\documentclass[twocolumn,reprint,superscriptaddress,amsmath,amssymb,aps,pra]{revtex4-2}
\usepackage{units}
\usepackage{graphicx,subfigure}
\usepackage{mathrsfs}
\usepackage{bm}% bold math
\usepackage{textcomp}
\usepackage{url}
\usepackage{dsfont} 
\usepackage{braket}
\usepackage[compact]{titlesec}
\usepackage[colorlinks=true,linkcolor=magenta,citecolor=blue]{hyperref}%
\usepackage{todonotes}
\usepackage{empheq}
\usepackage{multirow}
\begin{document}
\title{Goos-H\"{a}nchen Shift and Slow Light Enhancement in a Fixed Cavity: Bose-Einstein Condensate Bogoliubov Modes as Mechanical Oscillators}
\author{Ghaisud Din}
\affiliation{Ministry of Education Key Laboratory for Nonequilibrium Synthesis and Modulation of Condensed Matter, Shaanxi Province Key Laboratory of Quantum Information and Quantum Optoelectronic Devices, School of Physics, Xi’an Jiaotong University, Xi’an 710049, China}
\author{Fazal Badshah}
\affiliation{School  of Artificial Intelligence, Shiyan Key Laboratory of Quantum Information and Precision Optics, Hubei University of Automotive Technology, Shiyan 442002, China}
\author{Muqaddar Abbas}
\email{muqaddarabbas@xjtu.edu.cn}
\affiliation{Ministry of Education Key Laboratory for Nonequilibrium Synthesis and Modulation of Condensed Matter, Shaanxi Province Key Laboratory of Quantum Information and Quantum Optoelectronic Devices, School of Physics, Xi’an Jiaotong University, Xi’an 710049, China}
\author{Yunlong Wang}
\email{yunlong.wang@mail.xjtu.edu.cn}
\affiliation{Ministry of Education Key Laboratory for Nonequilibrium Synthesis and Modulation of Condensed Matter, Shaanxi Province Key Laboratory of Quantum Information and Quantum Optoelectronic Devices, School of Physics, Xi’an Jiaotong University, Xi’an 710049, China}
\author{Feiran Wang}
\email{feiran0325@xjtu.edu.cn}
\affiliation{School of Science, Xi’an Polytechnic University, Xi’an 710048, China}
\author{Pei Zhang}
\email{zhangpei@mail.ustc.edu.cn}
\affiliation{Ministry of Education Key Laboratory for Nonequilibrium Synthesis and Modulation of Condensed Matter, Shaanxi Province Key Laboratory of Quantum Information and Quantum Optoelectronic Devices, School of Physics, Xi’an Jiaotong University, Xi’an 710049, China}

\begin{abstract}
In this study, we explore the dynamics of slow and fast light propagation in a system consisting of a Bose-Einstein condensate (BEC) acting as a mechanical oscillator coupled to an optical parametric amplifier (OPA) within a fixed-mirror cavity. The system's response is investigated through a comprehensive analysis of the transmission spectrum, output probe field characteristics (real and imaginary components), group delay, and Goos-Hänchen shift (GHS). Our findings reveal that variations in the effective coupling strength and the OPA gain have a profound impact on the system's behavior. Specifically, as the OPA gain increases, a Fano-like resonance emerges, enhancing the transparency window and altering the dispersion, which in turn influences the group delay. The GHS is shown to be sensitive to both the incident angle and the BEC-cavity coupling strength. These results offer valuable insights into the intricate interplay between the probe field, the mechanical oscillator, and the amplified modes of the OPA, highlighting the role of these interactions in shaping the propagation of light in such systems.
\end{abstract}
\maketitle
\section{INTRODUCTION}
\label{Sec:intro}
The Goos-H\"{a}nchen shift (GHS) is a well-known optical phenomenon in which a beam of light, upon reflection from a boundary between two media with different refractive indices, experiences a lateral displacement along the interface. Initially theorized by Picht \cite{picht1929beitrag} and experimentally confirmed by Goos and Hänchen in a total internal reflection experiment using a glass slab \cite{goos1947new}. The theoretical foundations of the GHS were further developed by Artmann, who used the stationary phase method \cite{artmann1948berechnung}. Renard later proposed an alternative explanation based on the energy flux approach \cite{wang2013all}.

The GHS is significant in various optical applications, including high-precision optical sensors \cite{wang2008oscillating, hashimoto1989optical, hsue1985lateral}, beam splitting technologies \cite{song2012giant}, and temperature-dependent optical sensing \cite{chen2007optical}. The magnitude and direction of the shift depend on the properties of the interacting media, which can lead to either positive or negative GHS in different configurations. This effect has been widely studied in weakly absorbing materials \cite{wang2010lateral}, dielectric slabs with low absorption \cite{wang2005large, wang2010large}, optical gain media \cite{yan2007large}, negative refractive index materials \cite{shadrivov2003giant}, left-handed metamaterials \cite{wang2005large}, and photonic crystals \cite{felbacq2003goos}.

Moreover, Scully proposed a method to modify the susceptibility of a two-level atomic system by employing a coherently prepared ground-state doublet, allowing for active control over the GHS \cite{PhysRevLett.67.1855}. To further enhance tunability, researchers employed a classical coherent control field in a two-level atomic medium, enabling dynamic manipulation of the lateral shift \cite{PhysRevA.77.023811}. In a four-level atomic system embedded in an optical cavity, significant GHS enhancement was observed within the spectral hole-burning regime, both with and without Doppler broadening \cite{bacha2019implications}. Additionally, a $\Lambda$-type atomic medium was utilized to control GHS via electromagnetically induced transparency and amplification (EITA), demonstrating that adjusting the probe field frequency around EITA could dynamically switch the GHS between large positive and negative values through modulation of the collective phase of external fields \cite{deng2012enhancement, abbas2024goos}.

Recent progress in the field of light-matter interactions has paved the way for the generation of innovative quantum optical systems, offering superior control over their dynamics \cite{badshah2025coherent, zhou2024realization, badshah2024modulating, badshah2024investigation}.
Another captivating phenomenon that has captured significant attention is Bose-Einstein condensates (BECs), renowned for their remarkable quantum mechanical properties that distinguish them from other states of matter \cite{brennecke2008cavity}. Their ability to manipulate collective atomic motion has paved the way for controlling light propagation \cite{brennecke2008cavity, PhysRevA.83.055803}. In a dilute atomic gas, the achievement of BEC has led to substantial advances in quantum physics \cite{anderson1995observation, PhysRevLett.75.3969}. As systems where macroscopic quantum phenomena become accessible, BECs have provided insights into coherence, superfluidity, and quantum phase transitions \cite{RevModPhys.71.463, botelho2005quantum, aizenman2004bose}. In recent years, the integration of BECs into optical cavities has gained considerable interest due to the interplay between collective atomic motion and the quantized electromagnetic field \cite{RevModPhys.85.553, mekhov2007probing}, providing a rich platform for exploring fundamental and emergent phenomena such as self-organized supersolid phases \cite{baumann2010dicke}, superradiant phase transitions \cite{PhysRevA.91.021602}, and quantum optical nonlinearities \cite{PhysRevA.95.043601}.

Moreover, it has become increasingly relevant to view a BEC not only as a quantum field but also as a system with mechanical degrees of freedom. The collective excitations of the condensate, such as Bogoliubov modes or the center-of-mass motion, behave similarly to a mechanical oscillator \cite{PhysRevA.81.053833, nagy2009nonlinear, PhysRevA.64.033422}. These excitations couple to the intracavity field and are sensitive to photon number changes, allowing for light-induced control of the condensate dynamics.

This concept is supported by both theoretical and experimental studies that demonstrate the BEC's role as a mechanical element in hybrid quantum systems. For example, coupling between the BEC density and the intracavity field modifies both the atomic motion and the optical response of the system \cite{PhysRevLett.95.260401, brennecke2008cavity}.

In this study, we investigate the dynamics of laser propagation and the GHS shift within a system where a BEC functions as a mechanical oscillator coupled to optical cavity modes and an OPA inside the fixed cavity. By adjusting the coupling strength and OPA gain, we anticipate observing a Fano-like resonance that alters the dispersion and broadens the transparency window \cite{brennecke2007cavity, murch2008observation}. Moreover, we will examine the enhancement of slow and fast light propagation induced by the OPA's effect, resulting from the Fano resonance in the system. The study of slow light and GHS in BEC systems is vital for quantum information processing, optical communication, and precision measurement. Understanding these dynamics enhances the capabilities of quantum optical devices and opens avenues for new technologies based on slow light and light-matter interactions \cite{PhysRevA.84.055802}. Additionally, the GHS is influenced by the coupling strength between the BEC and the cavity, revealing a dependence on both the angle of incidence and the system's parameters.

\section{SYSTEM AND HAMILTONIAN}\label{section:MODEL} 
\begin{figure}
        \centering
        \includegraphics[width=1\linewidth]{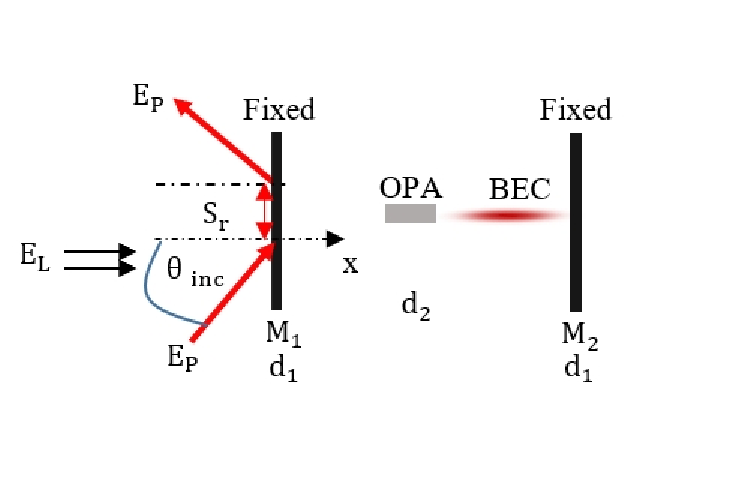}
        \caption{The schematic of the cavity system is shown, which includes a BEC of $N$ $^{87}\text{Rb}$ atoms and OPA. The red arrows represent the incoming and outgoing probe light from the left mirror $\text{M}_1$ with an incident angle $\theta_{\text{inc}}$. The black arrows indicate the control field, while $\text{S}_{\text{r}}$ denotes the shift in the reflected probe light. Both cavity mirrors, $\text{M}_1$ and $\text{M}_2$, are fixed in place.
}
        \label{figure1}
    \end{figure}

We consider a fixed cavity shown in Fig.\ref{figure1}, which contains a BEC of $N$ $^{87}\text{Rb}$ atoms along with an optical parametric amplifier (OPA). The cavity is  simultaneously driven by a control field with laser power $\lvert \text{E}_{\text{L}} \rvert$=$\sqrt{\frac{2\kappa \text{P}_{\text{L}}}{\hbar\omega_{\text{L}}}}$, with frequency $\omega_{\text{L}}$, and a weak probe field incident at an angle $\theta_{\text{inc}}$ from the left mirror $\text{M}_1$ having laser power  $\lvert \text{E}_{\text{P}} \rvert$=$\sqrt{\frac{2\kappa \text{P}_{\text{P}}}{\hbar\omega_{\text{P}}}}$ with frequency $\omega_{\text{P}}$. The probe light is reflected with a shift, denoted by $\text{S}_{\text{r}}$. So far the total Hamiltonian of the system containing BEC can be expressed as
\begin{equation}
  \mathscr H=  \int \text{d}\text{x} \Psi^\dagger(\text{x})[\frac{-\hbar^2}{2\text{m}}\frac{\text{d}^2}{\text{d}\text{x}^2}+\text{V}_{ext}(\text{x})+\hbar \text{U}_{0}\text{Cos}^2(\text{k}\text{x})\text{a}^\dagger \text{a}]\Psi(\text{x})\nonumber
\end{equation}
\begin{equation}
    +\hbar\omega_{\text{a}}\text{a}^\dagger \text{a}+\text{i}\hbar \text{G}(\text{a}^{\dagger^2}\text{e}^{\text{i}\theta}-\text{a}^2\text{e}^{-\text{i}\theta})+ \text{i}\hbar \text{E}_{L}(\text{a}^\dagger \text{e}^{-\text{i}\omega_L \text{t}}-\text{a}\text{e}^{\text{i}\omega_L \text{t}})\nonumber
\end{equation}
\begin{equation}
    + \text{i}\hbar \text{E}_{P}(\text{a}^\dagger e^{-\text{i}\omega_p \text{t}}-\text{a}e^{\text{i}\omega_P \text{t}})
\end{equation}
In the above Hamiltonian $\Psi^\dagger$, and $\text{a}^\dagger$ denote the creation operators of the atoms and the cavity photons respectively, where m shows the mass of the atom. $\text{V}_{\text{ext}}$ is the external potential, $\text{k}$ is the wave vector defined as $\text{k}=\frac{2\pi}{\lambda}$, and $\text{cos}(\text{k}\text{x})$ is the mode function. The atom experiences a maximum light shift in the cavity mode denoted by $\text{U}_{0}$ and defined as $\text{U}_{0}=\frac{\text{g}_{0}^2}{\Delta_{\text{a}}}$ where $g_{0}$ is the coupling strength between atom and photons, and $\Delta_\text{a}$ is the detuning of the atom frequency from the cavity field.
In the second and third line of the Hamiltonian in equation (1) $\omega_\text{a}$ is the frequency of the cavity, $\text{G}$ is the Gain of the OPA with phase $\theta$, $\text{E}_L$ and $\text{E}_P$ are the driving fields interacting with the cavity.
By applying the rotating wave approximation at the laser frequency $\omega_L$ and after the Bogoliubov approximation,
the Hamiltonian of the total system can be written as
\begin{equation}
   {\cal H}=\hbar \Delta_{\text{a}}\text{a}^\dagger \text{a}+\hbar\omega_{\text{m}}\text{b}^\dagger \text{b}+\hbar \text{g}_{bc}\text{a}^\dagger \text{a}(\text{b}+\text{b}^\dagger)+\text{i}\hbar\text{E}_{L}(\text{a}^\dagger-\text{a})
   \nonumber 
\end{equation}
\begin{equation}
   + \text{i}\hbar \text{E}_{P}(\text{a}^\dagger \text{e}^{-\text{i}\delta \text{t}}-\text{a}\text{e}^{\text{i}\delta \text{t}})+\text{i}\hbar \text{G}(\text{a}^{\dagger^2}\text{e}^{\text{i}\theta}-\text{a}^2\text{e}^{-\text{i}\theta})
\end{equation}

In the above Hamiltonian the first term denotes the free energy of the cavity with $\Delta_\text{a}=\omega_\text{a}+\frac{\text{U}_0 \text{N}}{2}-\omega_L$, and the annihilation(creation) operator $\text{a}(\text{a}^\dagger)$. The second term denotes the energy of the BEC with frequency $\omega_\text{m}=4\omega_{\text{rec}}$, where $\omega_{\text{rec}}=\frac{\hbar \text{k}^2}{2\text{m}}$ is the recoil frequency. The third term describes the interaction between the cavity and the Bogoliubov modes with annihilation(creation operator) $\text{b}(\text{b}^\dagger)$ where $\text{g}_{bc}$ denotes the coupling strength which is defined by $\text{g}_{bc}=\frac{\text{U}_0}{2}\sqrt{\frac{\text{N}}{2}}$. In the fifth term of the Hamiltonian in equation (2), $\delta$ denotes the control-probe detuning, defined as $\delta=\omega_\text{P}-\omega_\text{L}$.

By employing the Heisenberg equations of motion and incorporating both the damping and noise terms, we derive the quantum Langevin equations in the following form
\begin{eqnarray}
\Dot {\text{a}}&=&-(\text{i}\Delta_{\text{a}} + \kappa)\text{a} -\text{i} \text{g}_{bc}\text{a}(\text{b}+\text{b}^\dagger)+2\text{G}\text{e}^{\text{i}\theta}\text{a}^\dagger\nonumber \\[8pt]&&+\text{E}_{L}
+\text{E}_{P}\text{e}^{-\text{i}\delta \text{t}}+\sqrt{2\kappa}\text{a}_{in}\nonumber
\end{eqnarray}
\begin{eqnarray}
\Dot {\text{b}}&=&-(\text{i}\omega_{\text{m}} + \gamma_{\text{m}})\text{b} -\text{i} \text{g}_{bc}\text{a}^\dagger \text{a}+\sqrt{2\gamma_{\text{m}}}\xi
\end{eqnarray}
\begin{figure}
        \centering
        \includegraphics[width=1\linewidth]{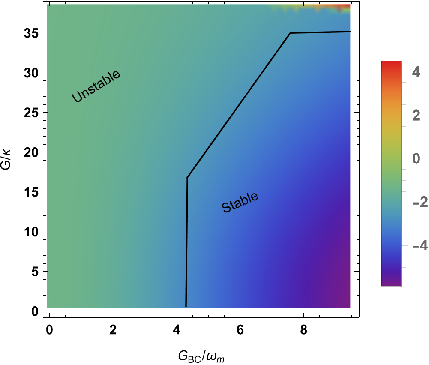}
        \caption{Density plot of the Real part of Eigenvalues of the characteristic equation of matrix A vs OPA G and coupling strength $\text{G}_{\text{BC}}$. Other parameters are $\omega_{\text{m}}/2\pi=15.2\text{kHz}$, $\gamma_{\text{m}}/2\pi=0.21\text{kHz}$, $\omega_{rec}/2\pi=3.8\text{kHz}$,$\lambda=780\text{nm}$,$\text{L}=1.25\times10^{-4}\text{m}$}
        \label{figure2}
    \end{figure}
 \begin{figure}
        \centering
        \includegraphics[width=1\linewidth]{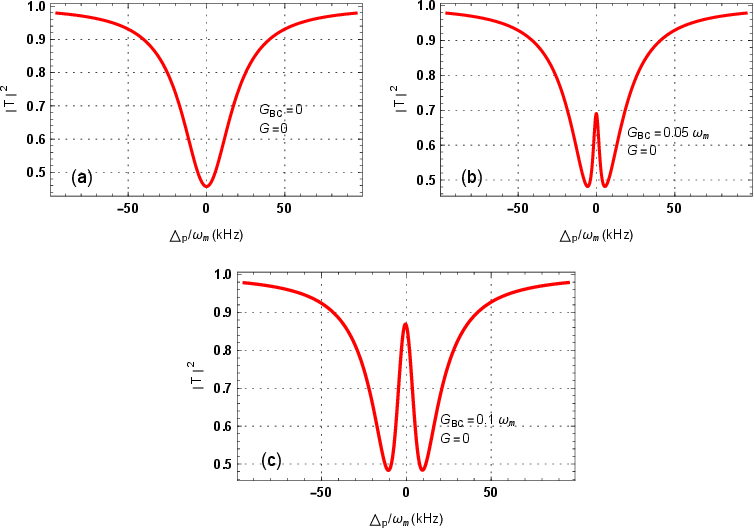}
        \caption{Transmission spectrum $\lvert \text{T} \rvert^2$, as a function of ${\bigtriangleup}_\mathbb{P}=\delta-\omega_{\text{m}}$ with parameters (a) $\text{G}_{\text{BC}}=0$, $\text{G}=0$,(b) $\text{G}_{\text{BC}}=0.05{\omega}_{\text{m}}$, $\text{G}=0$, (c) $\text{G}_{\text{BC}}=0.1{\omega}_{\text{m}}$, $\text{G}=0$. The other parameters are $\omega_{\text{m}}/2\pi=15.2\text{kHz}$, $\gamma_{\text{m}}/2\pi=0.21\text{kHz}$, $\omega_{rec}/2\pi=3.8\text{kHz}$,$\lambda=780\text{nm}$,$\text{L}=1.25\times10^{-4}\text{m}$.}
        \label{figure3}
    \end{figure}
     \begin{figure}
        \centering
        \includegraphics[width=1\linewidth]{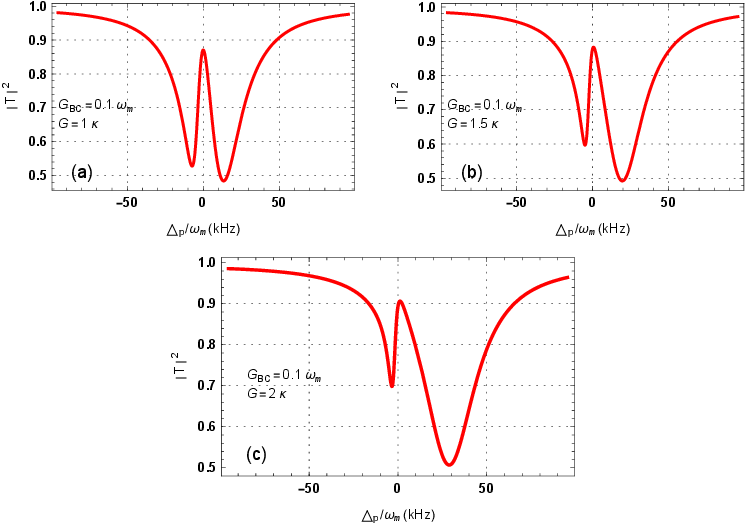}
        \caption{Transmission spectrum $\lvert \text{T} \rvert^2$,as a function of ${\bigtriangleup}_\mathbb{P}=\delta-\omega_{\text{m}}$  with parameters, $\text{G}_{\text{BC}}=0.1\omega_\text{m}$,$\theta=\pi/2$,$(a) \text{G}=1\kappa$,(b) $\text{G}=1.5\kappa$,(c) $\text{G}=2\kappa$. The other parameters are $\omega_{\text{m}}/2\pi=15.2\text{kHz}$, $\gamma_{\text{m}}/2\pi=0.21\text{kHz}$, $\omega_{rec}/2\pi=3.8\text{kHz}$,$\lambda=780\text{nm}$,$\text{L}=1.25\times10^{-4}\text{m}$.}
        \label{figure4}
    \end{figure}
    \begin{figure}
        \centering
        \includegraphics[width=1\linewidth]{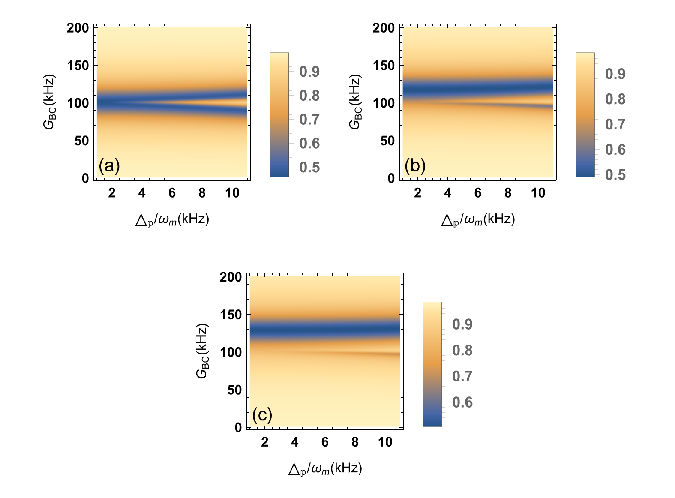}
        \caption{Density plot of the Transmission spectrum $\lvert \text{T} \rvert^2$,as a function of ${\bigtriangleup}_\mathbb{P}=\delta-\omega_{\text{m}}$ and coupling strength $\text{G}_{\text{BC}}$ with parameters,$\theta=\pi/2$,$(a) \text{G}=0$,(b) $\text{G}=1.5\kappa$,(c) $\text{G}=2\kappa$. The other parameters are $\omega_{\text{m}}/2\pi=15.2\text{kHz}$, $\gamma_{\text{m}}/2\pi=0.21\text{kHz}$, $\omega_{rec}/2\pi=3.8\text{kHz}$,$\lambda=780\text{nm}$,$\text{L}=1.25\times10^{-4}\text{m}$.}
        \label{figure5}
    \end{figure}
    \begin{figure}
        \centering
        \includegraphics[width=1\linewidth]{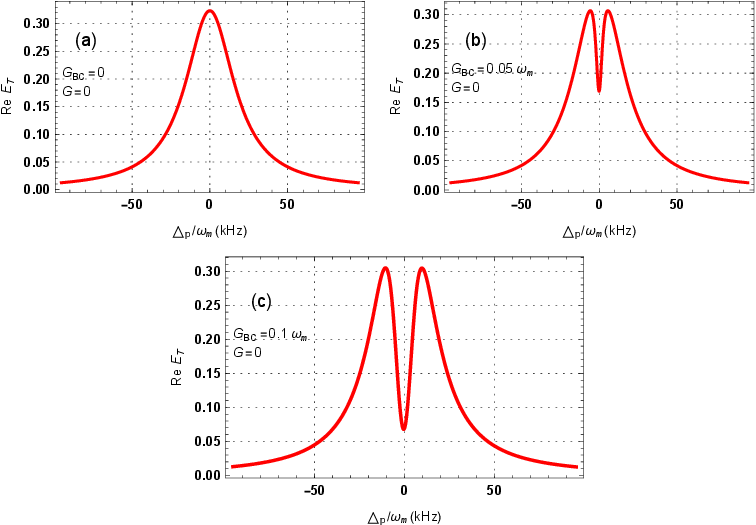}
        \caption{Real part of the output probe field spectrum $\text{Re}\text{E}_\text{T}$ as a function of ${\bigtriangleup}_\mathbb{P}=\delta-\omega_{\text{m}}$ with parameters, (a) $\text{G}_{\text{BC}}=0$, $\text{G}=0$,(b) $\text{G}_{\text{BC}}=0.05{\omega}_{\text{m}}$, $\text{G}=0$, (c) $\text{G}_{\text{BC}}=0.1{\omega}_{\text{m}}$, $\text{G}=0$. The other parameters are $\omega_{\text{m}}/2\pi=15.2\text{kHz}$, $\gamma_{\text{m}}/2\pi=0.21\text{kHz}$, $\omega_{rec}/2\pi=3.8\text{kHz}$,$\lambda=780\text{nm}$,$\text{L}=1.25\times10^{-4}\text{m}$.}
        \label{figure6}
    \end{figure}
     \begin{figure}
        \centering
        \includegraphics[width=1\linewidth]{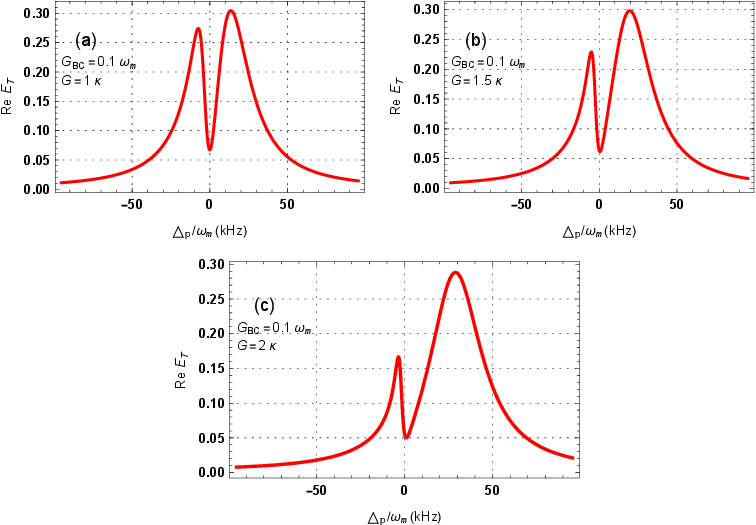}
        \caption{Real part of the output probe field spectrum $\text{Re}\text{E}_\text{T}$ as a function of ${\bigtriangleup}_\mathbb{P}=\delta-\omega_{\text{m}}$ with parameters, $\text{G}_{\text{BC}}=0.1\omega_\text{m}$,$\theta=\pi/2$,(a) $\text{G}=1\kappa$, (b) $\text{G}=1.5\kappa$, (c) $\text{G}=2\kappa$. The other parameters are $\omega_{\text{m}}/2\pi=15.2\text{kHz}$, $\gamma_{\text{m}}/2\pi=0.21\text{kHz}$, $\omega_{rec}/2\pi=3.8\text{kHz}$, $\lambda=780\text{nm}$, $\text{L}=1.25\times10^{-4}\text{m}$.}
        \label{figure7}
    \end{figure}
    \begin{figure}
        \centering
        \includegraphics[width=1\linewidth]{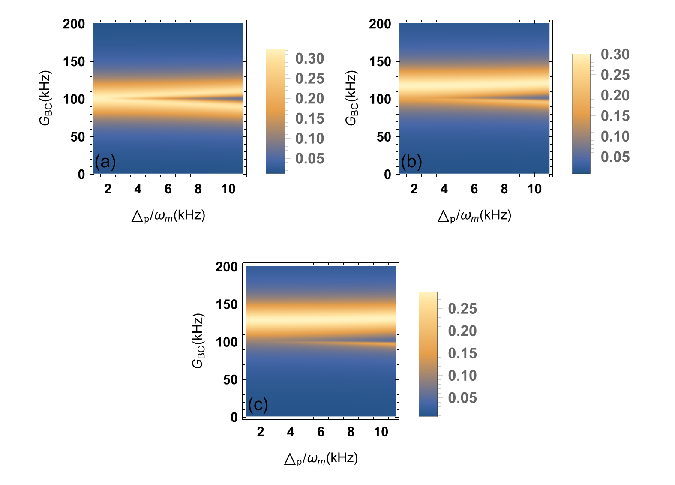}
        \caption{Density plot of the Real part of the output probe field spectrum $\text{Re}\text{E}_\text{T}$ as a function of ${\bigtriangleup}_\mathbb{P}=\delta-\omega_{\text{m}}$ and coupling strength $\text{G}_{\text{BC}}$ with parameters, $\theta=\pi/2$,$(a) \text{G}=0$, (b) $\text{G}=1.5\kappa$,(c) $\text{G}=2\kappa$. The other parameters are $\omega_{\text{m}}/2\pi=15.2\text{kHz}$, $\gamma_{\text{m}}/2\pi=0.21\text{kHz}$, $\omega_{rec}/2\pi=3.8\text{kHz}$, $\lambda=780\text{nm}$, $\text{L}=1.25\times10^{-4}\text{m}$.}
        \label{figure8}
    \end{figure}
    \begin{figure}
        \centering
        \includegraphics[width=1\linewidth]{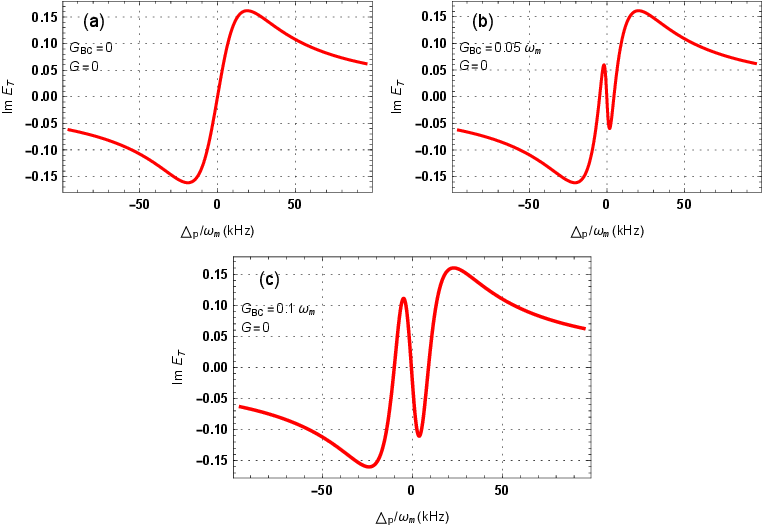}
        \caption{Imaginary part of the output probe field spectrum $\text{Im}(\text{E}_\text{T})$ as a function of ${\bigtriangleup}_\mathbb{P}=\delta-\omega_{\text{m}}$ with parameters, (a) $\text{G}_{\text{BC}}=0$, $\text{G}=0$, (b) $\text{G}_{\text{BC}}=0.05{\omega}_{\text{m}}$, $\text{G}=0$, (c) $\text{G}_{\text{BC}}=0.1{\omega}_{\text{m}}$, $\text{G}=0$. The other parameters are $\omega_{\text{m}}/2\pi=15.2\text{kHz}$, $\gamma_{\text{m}}/2\pi=0.21\text{kHz}$, $\omega_{rec}/2\pi=3.8\text{kHz}$, $\lambda=780\text{nm}$, $\text{L}=1.25\times10^{-4}\text{m}$.}
        \label{figure9}
    \end{figure}
    \begin{figure}
        \centering
        \includegraphics[width=1\linewidth]{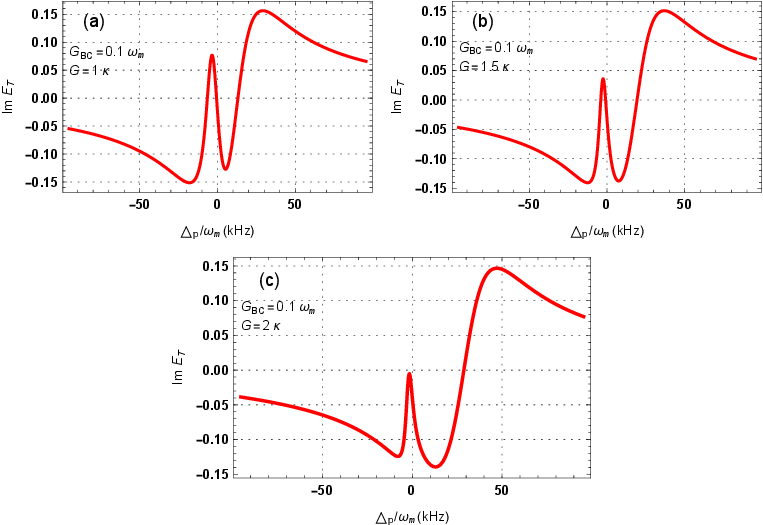}
        \caption{Imaginary part of the output probe field spectrum $\text{Im}(\text{E}_\text{T})$ as a function of ${\bigtriangleup}_\mathbb{P}=\delta-\omega_{\text{m}}$ with parameters, $\text{G}_{\text{BC}}=0.1\omega_\text{m}$,$\theta=\pi/2$, (a) $\text{G}=1\kappa$, (b) $\text{G}=1.5\kappa$, (c) $\text{G}=2\kappa$. The other parameters are $\omega_{\text{m}}/2\pi=15.2\text{kHz}$, $\gamma_{\text{m}}/2\pi=0.21\text{kHz}$, $\omega_{rec}/2\pi=3.8\text{kHz}$, $\lambda=780\text{nm}$, $\text{L}=1.25\times10^{-4}\text{m}$.}
        \label{figure10}
    \end{figure}
    \begin{figure}
        \centering
        \includegraphics[width=1\linewidth]{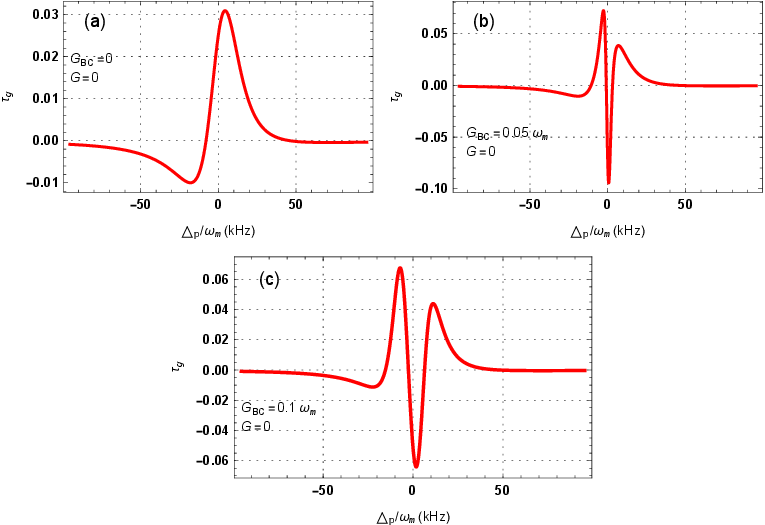}
        \caption{Group delay $\tau_{\text{g}}$ as a function of ${\bigtriangleup}_\mathbb{P}=\delta-\omega_{\text{m}}$ with parameters, (a) $\text{G}_{\text{BC}}=0$, $\text{G}=0$,(b) $\text{G}_{\text{BC}}=0.05{\omega}_{\text{m}}$, $\text{G}=0$, (c) $\text{G}_{\text{BC}}=0.1{\omega}_{\text{m}}$, $\text{G}=0$. The other parameters are $\omega_{\text{m}}/2\pi=15.2\text{kHz}$, $\gamma_{\text{m}}/2\pi=0.21\text{kHz}$, $\omega_{rec}/2\pi=3.8\text{kHz}$, $\lambda=780\text{nm}$, $\text{L}=1.25\times10^{-4}\text{m}$.}
        \label{figure11}
    \end{figure}
    \begin{figure}
        \centering
        \includegraphics[width=1\linewidth]{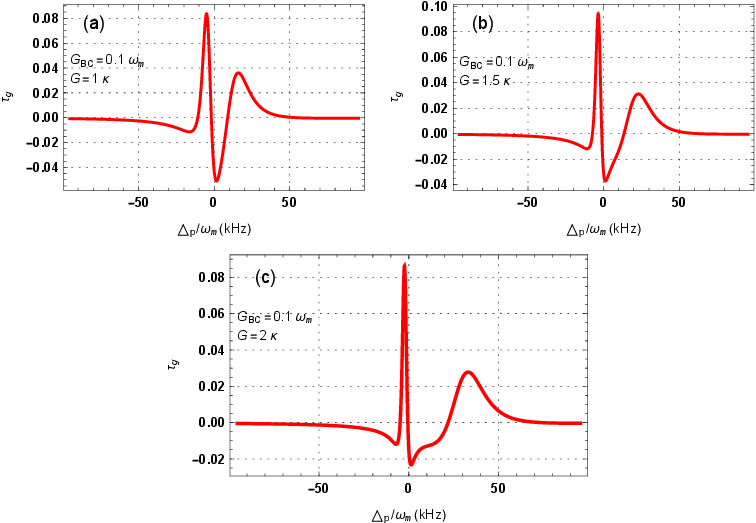}
        \caption{Group delay $\tau_{\text{g}}$ as a function of ${\bigtriangleup}_\mathbb{P}=\delta-\omega_{\text{m}}$ with parameters, $\text{G}_{\text{BC}}=0.1\omega_\text{m}$,$\theta=\pi/2$, (a) $\text{G}=1\kappa$, (b) $\text{G}=1.5\kappa$,(c) $\text{G}=2\kappa$. The other parameters are $\omega_{\text{m}}/2\pi=15.2\text{kHz}$, $\gamma_{\text{m}}/2\pi=0.21\text{kHz}$, $\omega_{rec}/2\pi=3.8\text{kHz}$, $\lambda=780\text{nm}$, $\text{L}=1.25\times10^{-4}\text{m}$.}
        \label{figure12}
    \end{figure}
    \begin{figure}
        \centering
        \includegraphics[width=1\linewidth]{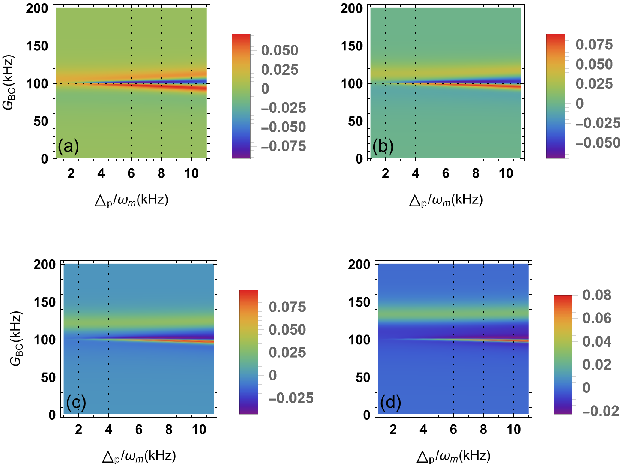}
        \caption{Density plot of the Group delay $\tau_{\text{g}}$ as a function of ${\bigtriangleup}_\mathbb{P}=\delta-\omega_{\text{m}}$ and coupling strength $\text{G}_{\text{BC}}$, with parameters $\theta=\pi/2$, (a) $\text{G}=0$,(b) $\text{G}=1\kappa$,(c) $\text{G}=1.2\kappa$, (d) $\text{G}=2\kappa$. The other parameters are $\omega_{\text{m}}/2\pi=15.2\text{kHz}$, $\gamma_{\text{m}}/2\pi=0.21\text{kHz}$, $\omega_{rec}/2\pi=3.8\text{kHz}$, $\lambda=780\text{nm}$, $\text{L}=1.25\times10^{-4}\text{m}$.}
        \label{figure13}
    \end{figure}
    \begin{figure}
        \centering
        \includegraphics[width=1\linewidth]{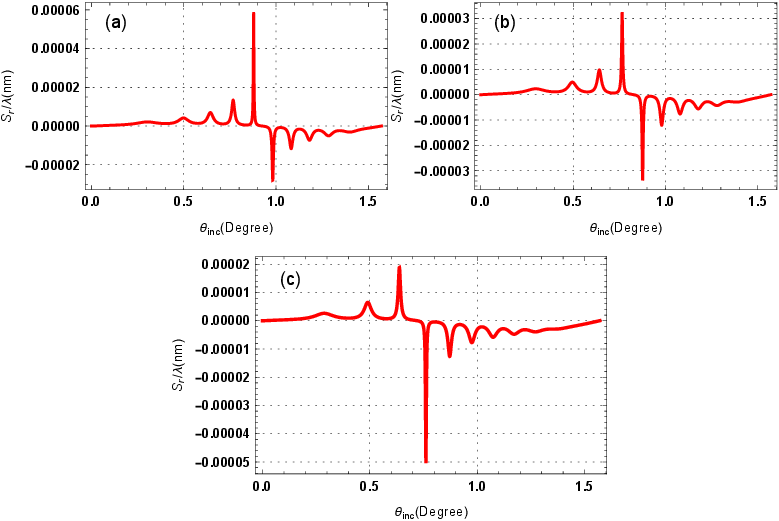}
        \caption{Goos-Hanchen shift $\text{S}_{r}/\lambda$ as a function of incident angle $\theta_\text{inc}$ for different OPA gain G, $\text{G}=0$(a), $\text{G}=1\kappa$ (b), $\text{G}=1.5\kappa$(c). While the coupling strength is fixed to  $\text{G}_{\text{BC}}=0.1\omega_\text{m}$.  The other parameters are $\omega_{\text{m}}/2\pi=15.2\text{kHz}$, $\gamma_{\text{m}}/2\pi=0.21\text{kHz}$, $\omega_{rec}/2\pi=3.8\text{kHz}$, $\lambda=780\text{nm}$, $\text{L}=1.25\times10^{-4}\text{m}$, $\epsilon_0=1$,$\epsilon_1=2.2$, mirror thickness $\text{d}_1=0.2\times10^{-6}\text{m}$, and $\text{d}_2=5\times10^{-6}\text{m}$.}
        \label{figure14}
    \end{figure}
     \begin{figure}
        \centering
        \includegraphics[width=1\linewidth]{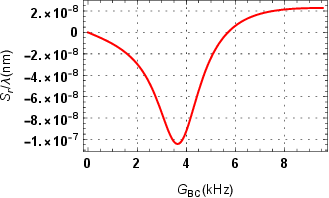}
        \caption{Goos-Hanchen shift $\text{S}_{r}/\lambda$ as a function of cavity BEC coupling strength $\text{G}_{\text{BC}}$  for an incident angle $\theta_\text{inc}=11^\circ$. While the gain of OPA is fixed to  $\text{G}=1.5\kappa$.  The other parameters are $\omega_{\text{m}}/2\pi=15.2\text{kHz}$, $\gamma_{\text{m}}/2\pi=0.21\text{kHz}$, $\omega_{rec}/2\pi=3.8\text{kHz}$, $\lambda=780\text{nm}$, $\text{L}=1.25\times10^{-4}\text{m}$, $\epsilon_0=1$,$\epsilon_1=2.2$, mirror thickness $\text{d}_1=0.2\times10^{-6}\text{m}$, and $\text{d}_2=5\times10^{-6}\text{m}$.}
        \label{figure15}
    \end{figure}
    \begin{figure}
        \centering
        \includegraphics[width=1\linewidth]{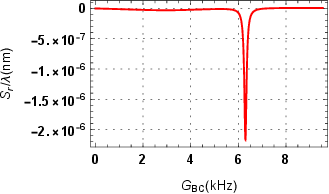}
        \caption{Goos-Hanchen shift $\text{S}_{r}/\lambda$ as a function of cavity BEC coupling strength $\text{G}_{\text{BC}}$  for an incident angle $\theta_\text{inc}=31^\circ$. While the gain of OPA is fixed to  $\text{G}=1.5\kappa$.  The other parameters are $\omega_{\text{m}}/2\pi=15.2\text{kHz}$, $\gamma_{\text{m}}/2\pi=0.21\text{kHz}$, $\omega_{rec}/2\pi=3.8\text{kHz}$, $\lambda=780\text{nm}$, $\text{L}=1.25\times10^{-4}\text{m}$, $\epsilon_0=1$,$\epsilon_1=2.2$, mirror thickness $\text{d}_1=0.2\times10^{-6}\text{m}$, and $\text{d}_2=5\times10^{-6}\text{m}$.}
        \label{figure16}
    \end{figure}
 Furthermore, the steady-state value plus a small fluctuation around that value is given by 
 $ \text{a}=\alpha+\delta \text{a}, \text{b}=\beta+\delta \text{b}$,
Using  these in equation (3) we get the following equations for the fluctuation operators

\begin{eqnarray}
\delta \Dot {\text{a}}&=&-(\text{i}\Delta + \kappa)\delta \text{a} -\text{i} \text{G}_{\text{BC}}(\delta \text{b}+\delta \text{b}^\dagger)+2\text{G}\text{e}^{\text{i}\theta}\delta \text{a}^\dagger\nonumber \\[8pt]&&+\text{E}_{P}\text{e}^{-\text{i}\delta \text{t}}\nonumber
\end{eqnarray}
\begin{eqnarray}
\delta \Dot {\text{b}}&=&-(\text{i}\omega_{\text{m}} + \gamma_{\text{m}})\delta \text{b} -\text{i} \text{G}_{\text{BC}}\delta \text{a}^\dagger -\text{i} \text{G}_{\text{BC}}^{*}\delta \text{a}
\end{eqnarray}

where $\text{G}_{\text{BC}}=\text{g}_{bc}\lvert\alpha\rvert$, $\Delta=\Delta_{\text{a}}-\text{g}_{bc}(\beta+\beta^*)$ and $\alpha$, $\beta$ are the steady-state solutions  which can be written as
\begin{eqnarray}
 \alpha=\frac{ \text{E}_{L}(2\text{G}\text{e}^{\text{i} \theta}+\kappa-\text{i}\Delta)}{-4\text{G}^2+\kappa^2+\Delta^2}
\end{eqnarray}
\begin{eqnarray}
 \beta=-\frac{\text{i} \text{g}_{bc}\lvert\alpha\rvert^2}{\text{i}\omega_{\text{m}}+\gamma_{\text{m}}}
\end{eqnarray}

In order to verify the stability of the system we define the cavity and mechanical quadrature with their corresponding input noise operator 
\begin{eqnarray}
   \text{x}&=&\frac{(\delta\text{a}^\dagger+\delta\text{a})}{\sqrt{2}},\text{y}=\frac{\text{i}(\delta\text{a}^\dagger-\delta\text{a})}{\sqrt{2}},\text{q}=\frac{(\delta\text{b}^\dagger+\delta\text{b})}{\sqrt{2}}\nonumber\\[8pt]&&\text{p}=\frac{\text{i}(\delta\text{b}^\dagger-\delta\text{b})}{\sqrt{2}},\text{x}^{\text{in}}=\frac{(\delta\text{a}^{\text{in}\dagger}+\delta\text{a}^{\text{in}})}{\sqrt{2}},\nonumber\\[8pt]&&\text{y}^{\text{in}}=\frac{\text{i}(\delta\text{a}^{\text{in}\dagger}-\delta\text{a}^{\text{in}})}{\sqrt{2}},\nonumber\\[8pt]&&\text{q}^{\text{in}}=\frac{(\delta\text{b}^{\text{in}\dagger}+\delta\text{b}^{\text{in}})}{\sqrt{2}},\text{p}^{\text{in}}=\frac{\text{i}(\delta\text{b}^{\text{in}\dagger}-\delta\text{b}^{\text{in}})}{\sqrt{2}}
\end{eqnarray}
Using equation 7 into equation 4 we can straightforwardly define the following matrix
\begin{equation}
    \text{A}=
\begin{pmatrix}
\text{x}_{11} & \text{x}_{12} & \text{x}_{13} & \text{x}_{14} \\
\text{x}_{21} & \text{x}_{22} & \text{x}_{23} & \text{x}_{24} \\
\text{x}_{31} & \text{x}_{32} & \text{x}_{33} & \text{x}_{34} \\
\text{x}_{41} & \text{x}_{42} & \text{x}_{43} & \text{x}_{44}
\end{pmatrix}
\end{equation}
where the elements of the matrix A can be expressed as
\begin{eqnarray}
    &\text{x}_{11}=&2\text{G}\text{cos}\theta-\kappa,\text{x}_{12}=\Delta+2\text{G}\text{sin}\theta,\text{x}_{13}=0,\nonumber\\[8pt]&&\text{x}_{14}=2\text{G}_{\text{BC}},\nonumber\text{x}_{21}=-\Delta+2\text{G}\text{sin}\theta,\nonumber\\[8pt]&&\text{x}_{22}=-\kappa+2\text{G}\text{cos}\theta,\text{x}_{23} =-2\text{G}_{\text{BC}},\text{x}_{24}=0\nonumber\\[8pt]&&\text{x}_{31}=0,\text{x}_{32}=\text{G}_{\text{BC}}+\text{G}_{\text{BC}}^*,\text{x}_{33}=-\gamma_{\text{m}},\nonumber\\[8pt]&&\text{x}_{34}=\omega_{\text{m}},\text{x}_{41}=-(\text{G}_{\text{BC}}+\text{G}_{\text{BC}}^*),\text{x}_{42}=0,\nonumber\\[8pt]&&\text{x}_{43}=-\omega_{\text{m}},\text{x}_{44}=-\gamma_{\text{m}}\nonumber
\end{eqnarray}
 Moreover to analyze the probe field response we drop the noise terms in equation (4) and  employ the following ansatz,
 \begin{equation}
     \delta \text{a}=\delta \text{a}_{-}\text{e}^{-\text{i}\delta \text{t}}+\delta \text{a}_{+}\text{e}^{\text{i}\delta \text{t}}
 \end{equation}
  \begin{equation}
     \delta \text{b}=\delta \text{b}_{-}\text{e}^{-\text{i}\delta \text{t}}+\delta \text{b}_{+}\text{e}^{\text{i}\delta \text{t}}
 \end{equation}
 Using the ansatz (9-10) in equation (4) we can obtain the output probe field response which is defined by $\text{a}_{-}$. To analyze the characteristics of the output probe field, we utilize the input-output relation given by
 \begin{equation}
\text{E}_{\text{out}}(\text{t})+\text{E}_{\text{P}}\text{e}^{-\iota\delta \text{t}}+\text{E}_{\text{L}}=\sqrt{2\kappa}\text{a}
\end{equation}
where
\begin{equation}
\text{E}_{\text{out}}(\text{t})=\text{E}_{\text{out}}^o+\text{E}_{\text{out}}^+\text{E}_{\text{P}}\text{e}^{-\iota\delta \text{t}}+\text{E}_{\text{out}}^-\text{E}_{\text{P}}\text{e}^{\iota\delta \text{t}}.
\end{equation}
By solving equations (11) and (12) we get the following expression
\begin{equation}
\text{E}_{\text{out}}^{+}=\frac{\sqrt{2\kappa} \text{a}_{{-}}}{\text{E}_{\text{P}}}-1
\end{equation}
And
\begin{equation}
\text{E}_{\text{out}}^{+}+1=\frac{\sqrt{2\kappa}\text{a}_{{-}}}{\text{E}_{\text{P}}}=\text{E}_{\text{T}},
\end{equation}
In equation (14) the output probe field response can be written as.
\begin{eqnarray}
 \text{a}_{-}=\frac{\cal A}{\cal B}
\end{eqnarray}
\begin{eqnarray}
 \cal A&=&-((\text{E}_{\text{P}}(2\text{i}\text{G}_{\text{BC}}^2\omega_{\text{m}}+(\gamma_\text{m}-\text{i}{\bigtriangleup}_\mathbb{P})(\gamma_\text{m}-\text{i}({\bigtriangleup}_\mathbb{P}+2\omega_\text{m}))\nonumber\\[8pt]&&(\kappa-\text{i}({\bigtriangleup}_\mathbb{P}+2\omega_\text{m}))))\nonumber
\end{eqnarray}
\begin{eqnarray}
 \cal B&=&(2\text{i}\text{G}_{\text{BC}}^2\kappa\omega_\text{m}+2\text{G}_{\text{BC}}^2{\bigtriangleup}_\mathbb{P}\omega_\text{m}+4\text{G}_{\text{BC}}^2\omega_{\text{m}}^2+4\text{e}^{2\text{i}\theta}\nonumber\\[8pt]&&\text{G}^2(\gamma_\text{m}-\text{i}{\bigtriangleup}_\mathbb{P})(\gamma_\text{m}-\text{i}({\bigtriangleup}_\mathbb{P}+2\omega_\text{m}))+(\kappa-\text{i}{\bigtriangleup}_\mathbb{P})\nonumber\\[8pt]&&(-2\text{i}\text{G}_{\text{BC}}^2\omega_\text{m}+(\gamma_\text{m}-\text{i}{\bigtriangleup}_\mathbb{P})(\text{i}\gamma_\text{m}+{\bigtriangleup}_\mathbb{P}+2\omega_\text{m})\nonumber\\[8pt]&&(\text{i}\kappa+{\bigtriangleup}_\mathbb{P}+2\omega_\text{m}))))\nonumber
\end{eqnarray}

Additionally by using equation (14) the transmission can be expressed as
\begin{equation}
     \lvert \text{T}\rvert^2=\lvert\frac{\sqrt{2\kappa} a_-} {\text{E}_{\text{P}}}-1\rvert^2
 \end{equation}
 while the group delay is given by
 \begin{equation}
      \tau_\text{g}=\frac{\partial\phi_{\text{T}}(\omega_\text{P})}  {\partial\omega_\text{P}}
 \end{equation}
Where $\phi_{\text{T}}$ is the phase of the transmitted probe field.
Next, we investigate the Goos-Hanchen shift, which is based on the stationary phase theory \cite{PhysRevLett.91.133903}, in this context the GHS for the reflected probe beam can be written as
\begin{eqnarray}
    \text{S}_r=-\frac{\lambda}{2\pi}\frac{\text{d}\phi_{\text{T}}}{\text{d}\theta}_{\text{inc}}
\end{eqnarray}
In equation (18) $\lambda$ denotes the wavelength of the incident probe light and $\phi_{\text{T}}$ is the phase of the transmitted beam $\text{T}(k_z,\omega_P)$ in which $k_z=\frac{2\pi\text{Sin}\theta}{\lambda}$. So far equation (18) can also be written in the following form
\begin{equation}
    \text{S}_r=-\frac{\lambda}{2\pi\lvert\text{T}\rvert^2}[\text{Re(T)}\frac{\text{d}}{\text{d}\theta}\text{Im(T)}+\text{Im(T)}\frac{\text{d}}{\text{d}\theta}\text{Re(T)}]
\end{equation}
here the transmission coefficient $\text{T}$ can be derived by using the transfer matrix method \cite{PhysRevA.77.023811,abbas2021enhancement} and can be written as
\begin{equation}
    \text{T}=\frac{2\text{Q}_0}{\text{Q}_0(\text{q}_{22}+\text{q}_{11})-(\text{Q}_{0}^2\text{q}_{12}+\text{q}_{21})}
\end{equation}
In equation (20) $\text{Q}_0=\sqrt{\epsilon_0-\text{sin}^2\theta}$ and $\text{q}_{ij}$ are the elements of the transfer matrix $\text{q}(\text{k}_{z},\omega_\text{P})=\text{m}_1(\text{k}_{z},\omega_P,\text{d}_1)\text{m}_2(\text{k}_{z},\omega_P,\text{d}_2)\text{m}_1(\text{k}_{z},\omega_P,\text{d}_1)$ with (ij=1,2) where $\text{d}_1$ is the thickness of the mirror $\text{M}_1$, $\text{M}_2$ and $\text{d}_2$ is related to the intracavity with the Bose-Einstein condensate and OPA. The elements $\text{m}_j$ are related to the probe field and can be expressed as 
\begin{equation}
    \text{m}_j(\text{k}_z,\omega_P,\text{d}_1)=\begin{bmatrix}
a_{11} & a_{12} \\
a_{21} & a_{22}
\end{bmatrix}
\end{equation}
with
\begin{eqnarray}
    &a_{11}&=\text{cos}(\text{k}^x_j\text{d}_j)\nonumber  \\[8pt]&& \nonumber  a_{12}=\text{i}\text{sin}(\text{k}^x_j\text{d}_j)\frac{\text{k}}{\text{k}^x_j},\\[8pt]&& \nonumber a_{21}=\text{cos}(\text{k}^x_j\text{d}_j), \\[8pt]&& \nonumber a_{22}=\text{i}\text{sin}(\text{k}^x_j\text{d}_j)\frac{\text{k}}{\text{k}^x_j}\nonumber 
\end{eqnarray}
where $\text{k}^x_j=(\frac{\omega_\text{P}}{c})\sqrt{\epsilon_j-\text{sin}^2\theta}$
\section{RESULTS AND DISCUSSION}\label{RESULTS}
We employ experimentally feasible parameters \cite{brennecke2007cavity,colombe2007strong} $\omega_{\text{m}}/2\pi=15.2\text{kHz}$, $\gamma_{\text{m}}/2\pi=0.21\text{kHz}$, $\omega_{rec}/2\pi=3.8\text{kHz}$, $\lambda=780\text{nm}$, $\text{L}=1.25\times10^{-4}\text{m}$. For the GHS we use additional parameters $\epsilon_0=1$, $\epsilon_1=2.2$, mirror thickness $\text{d}_1=0.2\times10^{-6}\text{m}$, $\text{d}_2=5\times10^{-6}\text{m}$.

Fig. \ref{figure2} density plot shows the real parts of the eigenvalues of the matrix A given by equation (8) of the system as functions of the OPA gain G and the coupling strength  $G_{\text{BC}}$. The values are color-coded to indicate different stability regimes. The blue region represents the stable domain, where all real parts of the eigenvalues are negative, implying the system naturally returns to its steady state after small disturbances. The warmer (Green) colors indicate the unstable region, where  eigenvalue has a positive real part, leading to exponential growth of perturbations. This visualization confirms how the stability of the system depends sensitively on both the OPA gain and the coupling strength, clearly separating the parameter space into stable and unstable zones.

Fig. \ref{figure3}(a-c) show how the probe transmission spectrum of the cavity BEC system evolves as the effective coupling strength between the cavity mode and the Bogoliubov mode $\text{G}_{\text{BC}}$ of the BEC is gradually increased while keeping the OPA gain zero (without any parametric amplification.). In Fig. \ref{figure3}(a), the coupling strength is set to zero, meaning there is no interaction of cavity modes with BEC. As a result, the cavity behaves like an empty Fabry–Pérot cavity, and the probe transmission displays a typical Lorentzian dip centered at the cavity resonance due to destructive interference from the cavity alone.

In Fig. \ref{figure3}(b), we introduce a small but non-zero coupling between the cavity field and the collective excitation mode of the BEC. Physically, this coupling arises from the dispersive interaction between the intra-cavity photons and the BEC density fluctuations. At this point, the system starts to exhibit weak transmission. A narrow transmission feature begins to appear near the frequency where the probe detuning matches the effective mechanical frequency $\omega_{\text{m}}$ of the BEC. This occurs because a fraction of the probe field is scattered by the BEC excitation and reenters the cavity in phase with the original probe field, partially canceling the destructive interference and leading to enhanced transmission.

Finally, in Fig.\ref{figure3}(c), the effective coupling $\text{G}_{\text{BC}}$ is further increased, strengthening the interaction between light and the BEC. This results in a clearer and sharper transmission peak in the spectrum. The interference becomes strongly destructive for the absorption pathway, while constructive interference enhances transmission at the mechanical resonance frequency. This hybridization between optical and atomic (mechanical) degrees of freedom modifies the probe field dynamics. Overall, these three figures demonstrate how the emergence and depth of the transmission depend sensitively on the cavity BEC coupling strength $\text{G}_{\text{BC}}$.

Next we demonstrate the evolution of the probe transmission spectrum as the gain parameter $\text{G}$ of the  OPA inside the cavity is gradually increased, while the effective coupling strength between the cavity field and the BEC is kept fixed. In Fig.\ref{figure4}(a) the OPA gain is increased slightly. The presence of the OPA modifies the intra-cavity field by introducing fluctuations and effectively enhances the interaction. As a result, the transmission spectrum becomes asymmetric and begins to exhibit a Fano-like profile. This asymmetry arises due to interference between a narrow transmission peak induced by the BEC and a modified light generated from the OPA.

In Fig.\ref{figure4}(b-c), the gain G is further increased (within the stability region shown in Fig.\ref{figure2}), and the impact of the parametric process becomes stronger. The interference effects become more pronounced, and the transmission peak evolves into a sharp and asymmetric Fano resonance. This transition reflects the increasing role of quantum interference. These results demonstrate that the OPA gain control light transmission in our system where the BEC serve as a mechanical mode(no physical mirror), enabling tunable interference features even when the cavity BEC coupling remains constant.

Further in Fig.\ref{figure5}(a-c) we present density plots of the probe transmission as a function of probe detuning ${\bigtriangleup}_\mathbb{P}=\delta-\omega_{\text{m}}$ (horizontal axis) and the effective coupling strength between the cavity field and the BEC $\text{G}_{\text{BC}}$ (vertical axis), for increasing values of the OPA gain $\text{G}$. In each plot, brighter regions indicate higher transmission. These figures reveal how the interplay between the BEC cavity interaction and the OPA shapes the transmission spectrum.

In Fig.\ref{figure5}(a), the OPA gain is set to zero, so the system operates as a conventional cavity BEC setup. At zero OPA gain, the transmission exhibits a standard peak profile.

In Fig.\ref{figure5}(b), a moderate OPA gain $\text{G}=1.5\kappa$ is introduced. The intra-cavity field is influenced by the parametric amplification process, which modifies both the amplitude and phase of the field. This causes the transmission to develop a slight asymmetry across the de-tuning axis, manifesting as a Fano-like resonance. Moreover, the transmission becomes sharper and more sensitive to changes in coupling strength, indicating enhanced interaction due to the effects introduced by the OPA.

Next, in Fig.\ref{figure5}(c), the OPA gain is further increased, significantly amplifying the fluctuations within the cavity. The transmission spectrum now shows a strongly asymmetric and distorted transmission, which bends and narrows along the detuning axis. The Fano resonance becomes more pronounced and shifts slightly due to the OPA-altered effective cavity detuning and linewidth. Additionally, the enhanced intracavity photon number modifies the system’s susceptibility, leading to stronger dispersion and greater contrast in the density map. This figure clearly shows that the OPA gain acts as a powerful control knob, tuning both the position and shape of the transmission features  even though the cavity–BEC coupling is fixed along each horizontal line.

So far Fig.\ref{figure6}(a-c) display the real part of the output probe field as a function of probe detuning, for increasing values of the effective coupling strength $\text{G}_{\text{BC}}$ between the cavity and the BEC. In this context, the real part is associated with the absorption spectrum of the probe field, and thus directly reveals how energy is transferred from the probe into the cavity system. Throughout all three figures, the optical parametric amplifier OPA gain is kept at zero.

In Fig.\ref{figure6}(a), the coupling strength $\text{G}_{\text{BC}}$ is zero, and the system behaves like an empty optical cavity. The probe experiences maximum absorption at zero detuning, leading to a symmetric Lorentzian absorption peak, the typical signature of a cavity with no interference effects.

In Fig.\ref{figure6}(b), the coupling between the cavity and the BEC is turned on. The BEC now acts as an effective mechanical oscillator, interacting with the intracavity photons through radiation pressure-like coupling. This leads to interference between the directly transmitted probe and the probe field scattered by the BEC. As a result, a narrow transparency window begins to emerge at the resonance detuning, suppressing absorption at that frequency and forming a characteristic dip.

In Fig.\ref{figure6}(c), the coupling strength is increased further, enhancing the interaction. The destructive interference becomes stronger and more efficient, giving rise to a well-defined and sharper transparency window, and is the hallmark of coherent energy exchange between the probe field and the hybridized cavity BEC system. These results clearly illustrate how increasing the cavity BEC coupling strength transforms the system’s absorption profile and leads to the formation of quantum interference-induced transparency.

Furthermore, Fig.\ref{figure7}(a-c) present the real part of the output probe field as a function of probe detuning, for increasing values of the optical parametric amplifier OPA gain G. In this case, the effective coupling strength between the cavity and the BEC is held fixed at $\text{G}_{\text{BC}}=0.1\omega_{\text{m}}$, such that a transparency window is already present in the absence of the OPA.

In Fig.\ref{figure7}(a), the OPA gain is increased moderately. As a result, the absorption profile becomes asymmetric, and the transparency window begins to resemble a Fano-like shape. This asymmetry stems from interference between the  probe and the vacuum field introduced by the OPA, which modifies the phase relationship between the two optical pathways.

In Fig.\ref{figure7}(b-c), a stronger OPA gain is applied leading to a sharper and more distorted transparency window with enhanced contrast and steeper dispersion near resonance. The increased intracavity photon number and induced modification of the cavity response deepen and narrow the transparency dip. Additionally, the transparency window may shift slightly due to the effective modification of the cavity detuning and linewidth by the OPA. These changes signal enhanced quantum interference and point to the potential for tuning absorption features using parametric control. Together, these figures demonstrate that increasing the OPA gain allows for precise engineering of the probe absorption spectrum.

Next Fig.\ref{figure8}(a-c) present density plots of the real part of the output probe field as a function of the probe detuning (horizontal axis) and the effective coupling strength between the cavity and the BEC (vertical axis). Across the three figures, the OPA gain G is progressively increased, revealing its impact on the formation and shape of the transparency window.

In Fig.\ref{figure8}(a), where the OPA gain is zero, the system behaves as a standard cavity BEC hybrid system. We observe a narrow transparency window emerges within the broader absorption band. This window widens and becomes more pronounced with stronger coupling, reflecting enhanced destructive interference between the direct and BEC scattered probe fields.

In Fig.\ref{figure8}(b), an OPA gain is introduced. The intracavity field begins to altering its phase and intensity fluctuations. This causes a visible deformation in the transparency window, the absorption profile becomes asymmetric, and the previously symmetric dip now resembles a Fano-like resonance.

In Fig.\ref{figure8}(c), with a higher OPA gain, the effect becomes more dramatic. The transparency window becomes even narrower, deeper, and strongly asymmetric. The enhanced quantum interference  driven by the interplay between light and BEC-cavity coupling leads to a highly tunable and sensitive absorption spectrum. These density plots demonstrate that increasing the OPA gain provides a powerful degree of control over the probe absorption, allowing for the reshaping and enhancement of transparency windows even when the cavity BEC coupling is not maximized.

Furthermore, Fig.\ref{figure9}(a-c) show the imaginary part of the output probe field plotted against probe detuning for increasing values of the effective coupling strength between the cavity and the BEC. In these plots, the OPA gain is kept fixed at zero, while the BEC cavity interaction strength is progressively enhanced from figure Fig.\ref{figure9}(a) to Fig.\ref{figure9}(c). The imaginary part represents the dispersive response of the system, which provides insight into the phase shift experienced by the probe and is directly related to group velocity and light delay within the cavity.

In Fig.\ref{figure9}(a), where the coupling strength is zero, the system acts as a standard cavity with minimal back action from the BEC. The dispersive curve is a smooth and symmetric S-shaped profile centered at zero detuning  characteristic of a typical resonant optical cavity. The slope around resonance is moderate, suggesting normal dispersion and minimal group delay.

In Fig.\ref{figure9}(b), with a moderate BEC cavity coupling strength $\text{G}_{\text{BC}}=0.05\omega_{\text{m}}$, the dispersive profile begins to steepen near resonance. This is due to the coherent energy exchange between the cavity photons and the collective excitation of the BEC. A sharp transition in the imaginary part appears around the transparency point, where phase variation becomes more abrupt. This steep slope corresponds to a large group delay.

In Fig.\ref{figure9}(c), the coupling strength is further increased, strongly enhancing the interaction. The result is a significantly sharper and more nonlinear dispersion curve. Near resonance, the imaginary part changes very rapidly with detuning, highlighting strong phase modulation effects. This steep phase response  supports enhanced slow light. Together, these three figures demonstrate that increasing the cavity BEC coupling strength strongly enhances the system's dispersive properties, leading to steeper phase variations and greater tunability of optical delay.

Similarly, Fig.\ref{figure19}(a-c) illustrate the imaginary part of the output probe field as a function of probe detuning for increasing values of the optical parametric amplifier OPA gain G, while the coupling strength between the cavity and the BEC is kept fixed.

In Fig.\ref{figure10}(a), the OPA gain is increased to a moderate value. The parametric amplification begins to influence the intracavity field, and modifying the interference conditions between the probe and the scattered field from the BEC. As a result, the dispersive profile becomes asymmetric, with a sharper transition near the transparency point, which introduces non-linear phase shifts, making the system more sensitive to detuning and enhancing group delay effects.

In Fig.\ref{figure10}(b-c), the OPA gain is further increased. The OPA significantly amplifies the intracavity field fluctuations, leading to a highly asymmetric and distorted dispersion profile. The transparency window becomes sharper and more pronounced, accompanied by steeper phase changes near resonance This behavior results from the OPA ability to modify the phase properties of the probe field, leading to enhanced slow light phenomena and greater tunability of the system's dispersive characteristics.

Next, Fig.\ref{figure11}(a-c) we demonstrate the group delay of the output probe field as a function of probe detuning for increasing values of the effective coupling strength between the cavity and the BEC. The OPA gain is fixed in these plots, and the BEC cavity coupling strength is varied across the three figures.

In Fig.\ref{figure11}(a), where the cavity BEC coupling strength is minimal, the group delay exhibits a relatively small variation across detuning. The delay remains fairly constant, and the system behaves as a standard optical cavity. The absence of strong interactions between the cavity photons and the BEC leads to a weak, nearly flat response, with no significant slowing of the probe light.

In Fig.\ref{figure11}(b-c), the coupling strength is increased, which enhances the interaction between the cavity and the collective excitations of the BEC. As a result, the group delay becomes more pronounced near the resonance detuning. The increased coupling strengthens the interaction, leading to more significant delays. It can be seen that slow-fast light appears at two different regions. The slow light appear where the group delay $\tau_{\text{g}}>0$ and fast light appears where the slope is negative i.e. $\tau_{\text{g}}<0$.

Furthermore, Fig.\ref{figure12}(a-c) show the group delay of the output probe field as a function of probe detuning for increasing values of the OPA gain G, while the effective coupling strength between the cavity and the BEC is held fixed.

In Fig.\ref{figure12}(a), when the OPA gain is increased to a moderate value, the group delay becomes more pronounced and exhibits greater asymmetry around resonance. The effect introduced by the OPA amplifies the interaction between the probe field and the BEC, leading to a steeper phase shift and asymmetric transparency window. The group delay grows more rapidly near the transparency point, and the asymmetry becomes more noticeable, enhancing the slow light effect. The increased asymmetry leads to a stronger dispersive response, causing the probe light to slow down more significantly as it passes through the cavity.

In Fig.\ref{figure12}(b-c), where the OPA gain is further increased, the asymmetry in the group delay becomes even more pronounced. This enhanced asymmetry leads to a more pronounced slow light effect, as the probe field is delayed more significantly at resonance. The increased OPA gain strengthens the non linear phase shift. As a result, the system exhibits strong slow-light behavior due to the asymmetry and effects introduced by the OPA, with the group delay becoming more sensitive to detuning and exhibiting larger delays.

Next, Fig.\ref{figure13}(a-d) display the group delay of the output probe field in a density plot format, where the x-axis corresponds to the probe detuning, the y-axis corresponds to the effective coupling strength between the cavity and the BEC, and the color intensity represents the magnitude of the group delay. The OPA gain G is varied across the foure figures, illustrating how the group delay evolves with increasing OPA gain.

In Fig.\ref{figure13}(a), where the OPA gain is set to $\text{G} = 0$, the density plot shows a relatively uniform distribution of group delay values. The group delay increases slightly near the transparency window, but there is no significant asymmetry. The transparency window is broad, and the group delay is moderate, reflecting the standard behavior of a cavity-BEC system without any parametric effects.

In Fig.\ref{figure13}(b), with the OPA gain set to $\text{G} = 1\kappa$, the density plot shows enhanced group delay compared to Fig.\ref{figure12}(a). The plot becomes asymmetric, with the intensity of group delay increasing more sharply near resonance detuning. The transparency window is now narrower, and the group delay exhibits stronger non-linear phase shifts due to the  effect of the OPA. This leads to a narrower and more intense transparency region in the density plot, indicating slow light effect.

In Fig.\ref{figure13}(c), with the OPA gain increased to $\text{G} = 1.5\kappa$, the group delay becomes even more pronounced and exhibits a steeper intensity gradient near resonance. The transparency window becomes narrower, and the asymmetry is even more prominent. The effects introduced by the OPA enhances the dispersive response further, leading to larger group delays.

In Fig.\ref{figure13}(d), where the OPA gain is set to $\text{G} = 2\kappa$, the group delay exhibits a stronger asymmetry compared to the previous figures. The transparency window becomes asymmetric, and the group delay increases dramatically. The effect is now at its maximum, leading to a highly dispersive system with a pronounced slow light effect.

Moreover, using Equation (18), we plot the GHS,  $S_{\text{r}} / \lambda$, as a function of the incident angle  $\theta_{\text{inc}}$ for different OPA gain, while setting the $G_{\text{BC}}$ fixed at $G_{\text{BC}}=0.1\omega{\text{m}}$ The GHS for  ${\text{G}}=0$ is represented by  Fig.\ref{figure14}(a), showing the shift at different incident angles. When the OPA gain under stability regions increases to ${\text{G}} =1\kappa$, the shift becomes more pronounced (negative), as indicated by Fig.\ref{figure14}(b). Further increasing the OPA gian to ${\text{G}}= 1.5\kappa$, results in an even larger (negative) shift, represented by Fig.\ref{figure14}(c).  

Furthermore, we plot the GHS, $S_{\text{r}} / \lambda$, as a function of the coupling strength $G_{\text{BC}}$ for different incident angles. In Fig.\ref{figure15} and Fig.\ref{figure16}, the shift is shown for an incident angle of $\theta_{\text{inc}} = 11^\circ$ and $\theta_{\text{inc}} = 31^\circ$.  
\section{CONCLUSION}\label{section:Conclusion} 
This study investigate the impact of a BEC serving as a mechanical oscillator in an optomechanical system, coupled with the effects of an OPA placed within the cavity. By varying the coupling strength $G_{\text{BC}}$ and the OPA gain, we observe distinct shifts in the transmission spectrum, the output probe field, and the group delay. Notably, the appearance of a Fano-like resonance with increasing OPA gain leads to asymmetric peaks and dips in the transmission spectrum, as well as sharper features in the real part of the output probe field. These effects are accompanied by enhanced slow light propagation and a pronounced shift in the group delay. Additionally, the study of the GHS as a function of incident angle reveals a strong dependence on the coupling strength, with the GHS showing notable enhancement at grazing incidence and vanishing at normal incidence. Overall, the findings of this work provide valuable insights into the control of light-matter interactions in fixed cavity systems, highlighting the potential for manipulating slow light and dispersion effects through careful tuning of system parameters.

\textbf{Data Availability Statement} This manuscript has associated data in a data repository. All data included in this paper are available upon request by contacting with the corresponding author.
\section*{Acknowledgements}
 This work was supported by the National Natural Science Foundation of China (Grant No. 12174301), the Young Investigator (Grant No. 12305027), the Natural Science Basic Research Program of Shaanxi (Program No. 2023-JC-JQ-01), and the Fundamental Research Funds for the Central Universities.
\bibliographystyle{apsrev4-2}
\bibliography{biblio.bib}
\end{document}